\documentclass[twocolumn,showpacs,preprintnumbers,amsmath,amssymb,superscriptaddress,pra,aps,10pt,nofootinbib,floatfix]{revtex4-1}

\usepackage[english]{babel}
\usepackage{graphicx}
\usepackage{dcolumn}
\usepackage{bm}
\usepackage{color}
\usepackage{textcomp}
\usepackage[lofdepth,lotdepth,caption=false]{subfig} 
\usepackage{textgreek}
\usepackage{multirow} 
\usepackage[utf8]{inputenc} 

\newcommand{\balpha}{\boldsymbol\alpha}
\newcommand{\bepsilon}{\boldsymbol\epsilon}

\newcommand{\ssBA}{\scriptscriptstyle BA}
\newcommand{\sss}{\scriptscriptstyle}

\newcommand{\removeit}[1]{}

\newcolumntype{1}{>{\cellcolor{tablebgI}}}
\newcolumntype{2}{>{\cellcolor{tablebgII}}}

\def\fig_width{3. in} 
\makeatletter
\newcommand{\thickhline}{%
    \noalign {\ifnum 0=`}\fi \hrule height 1pt
    \futurelet \reserved@a \@xhline
}
\makeatother
\begin{document}

\title{Investigation of ac-Stark shifts in excited states of dysprosium relevant to testing fundamental symmetries}
\author{C.T.M. Weber}
\email{ChristianTMWeber@gmail.com}
\affiliation{Technische Universit\"{a}t Berlin, 10623 Berlin, Germany}
\affiliation{Department of Physics, University of California at Berkeley, Berkeley, CA 94720-7300, USA}

\author{N. Leefer}
\email{naleefer@berkeley.edu}
\affiliation{Department of Physics, University of California at Berkeley, Berkeley, CA 94720-7300, USA}

\author{D. Budker}
\email{budker@berkeley.edu}
\affiliation{Department of Physics, University of California at Berkeley, Berkeley, CA 94720-7300, USA}
\affiliation{Nuclear Science Division, Lawrence Berkeley National Laboratory, Berkeley, CA 94720, USA}

\date{\today}



\begin{abstract}
We report on measurements of the differential polarizability between the nearly degenerate, opposite parity states in atomic dysprosium at 19797.96 cm$^{-1}$, and the differential blackbody radiation induced Stark shift of these states. The differential scalar and tensor polarizabilities due to additional states were measured for the $|M| = 7,\dots,10$ sublevels in $^{164}$Dy and $^{162}$Dy and determined to be $\overline{\balpha}_{\sss BA}^{(0)} = 180\,(45)_\text{stat}\,(8)_\text{sys}$ $\text{mHz}/\left( \mathrm{V}/\mathrm{cm}\right)^2$ and $\overline{\balpha}_{\sss BA}^{(2)} =  -163\,(65)_\text{stat}\,(5)_\text{sys}$ $\text{mHz}/\left(\mathrm{V}/\mathrm{cm}\right)^2$, respectively. The average blackbody radiation induced Stark shift of the Zeeman spectrum was measured around 300 K and found to be $-34(4)$~mHz/K and $+29(4)$~mHz/K for $^{164}$Dy and $^{162}$Dy, respectively. We conclude that ac-Stark related systematics will not limit a search for variation of the fine-structure constant, using dysprosium, down to the level of $|\dot{\alpha}/\alpha|=2.6\times10^{-17}$~yr$^{-1}$, for two measurements of the transition frequency one year apart.
\end{abstract}


\maketitle

\section{Introduction} 
\noindent
Some of the most precise tests of fundamental physics are clock-comparison experiments, where the frequencies of atomic or molecular transitions are compared against each other~\cite{Rosenband.2008,Leefer.2013}. The dependence of these frequencies on the parameters of atomic theory allows such comparisons to place stringent bounds on new physics, such as possible variation of fundamental constants~\cite{Cingoz.2007,Ferrell.2007,Guena.2012} or breaking of Lorentz symmetry and violation of the Einstein Equivalence Principle~\cite{HohenseeM.A..2013,Altschul.2010}. 

Interpreting these experiment's results requires atomic structure calculations that predict the dependence of transition frequencies on fundamental constants~\cite{Dzuba.2008}, e.g. the fine-structure constant $\alpha$, or various forms of Lorentz symmetry violation~\cite{Kostelecky1999a}. Since it is not possible to systematically vary fundamental constants or break Lorentz symmetry to verify these calculations, it is important to measure other parameters of atomic systems to test the validity of these calculations, such as relative energy splittings, state lifetimes, matrix elements~\cite{Budker.1994,Dzuba.2010}, and polarizabilities~\cite{Amini.2003,Ulzega2006}. The measurement of polarizabilities is also critically important for the evaluation of systematic errors due to Stark shifts in the measurement of transition frequencies ~\cite{Itano.1982,Levi.2004}.

Our group performs radio-frequency spectroscopy of an electric-dipole transition between nearly-degenerate, excited states in dysprosium (Dy) to search for variation of $\alpha$~\cite{Nguyen.2004,Cingoz.2007,Ferrell.2007,Leefer.2013} and constrain possible violation of Lorentz symmetry within the framework of the Standard Model Extension~\cite{HohenseeM.A..2013}. In this paper we report measurements of the differential electric-dipole polarizabilities~\cite{Bonin1994} of these states in Dy. These measurements are used to assess the current systematic limit due to ac-Stark shifts of the atomic levels in the search for variation of $\alpha$, and are an important check of the theory used in atomic structure calculations for heavy atoms with complex spectra.

\subsection*{Dysprosium} 
\begin{figure}[t]
 \includegraphics[width=1\columnwidth]{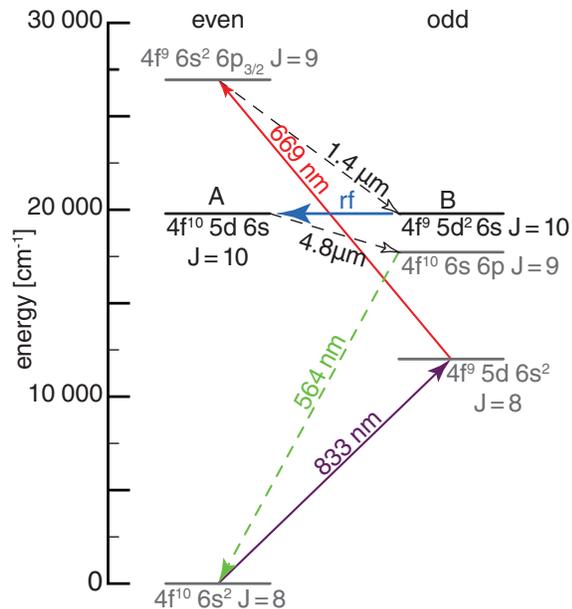}
\caption{Partial energy levels diagram of dysprosium. Solid (dashed) arrows indicate induced (spontaneous) transitions.}
\label{fig:popscheme}
\end{figure}

\noindent 
Dysprosium is a rare-earth element with nuclear charge $Z=66$ and seven stable isotopes with masses A = 156, 158, 160, 161, 162, 163, and 164. The energy level structure of Dy includes a pair of opposite parity, nearly-degenerate excited states, labeled $A$ and $B$ for the even and odd parity states, respectively~\cite{Budker.1994} (Fig.~\ref{fig:popscheme}). 
Prior to the studies of variation of $\alpha$ and Lorentz symmetry violation, this system was used to study atomic parity violation~\cite{Nguyen.1997}. In addition, the large ground-state magnetic moment of dysprosium ($\mu \approx 10\,\mu_B$) and recent advances in laser cooling~\cite{Leefer.2010}, trapping~\cite{Lu.2010}, and condensation~\cite{Lu.2011,Lu.2012} make Dy interesting in the context of quantum information processing~\cite{Derevianko.2004} and studies of exotic quantum phases of matter~\cite{Fregoso.2009}. 

The splitting between states $A$ and $B$ is sensitive to breaking of Lorentz symmetry due to the large difference in electron kinetic energies between the two states~\cite{HohenseeM.A..2013}, and the sensitivity to variation of $\alpha$ is given by large relativistic corrections to the electron energies~\cite{Dzuba.2008}. The small, isotope dependent splitting ($\approx 3 - 2000$~MHz) greatly relaxes the requirements on the long-term stability of a reference frequency~\cite{Nguyen.2004}. Traditional clock-comparison experiments must reach extraordinary levels of stability to place comparable bounds on variation of $\alpha$~\cite{Guena.2012,Rosenband.2008}. Reference~\cite{JeanPhilippeUzan.2011} gives a detailed overview of the theory and experiments concerning variation of $\alpha$ and other fundamental constants, and reference~\cite{DavidMattingly.2005} summarizes the theory and experiments concerning tests of Lorentz symmetry.
%

From here on we limit ourselves to the measurement of the differential polarizabilities of states A and B and the implications to the constraints of variation of $\alpha$. Such constraints are obtained by measuring the transition frequencies, $\nu_{\sss BA} = (\epsilon_{\sss B} -\epsilon_{\sss A})/h$, in $^{164}$Dy (+753.5 MHz) and $^{162}$Dy (-234.7 MHz) over several years. The absolute value of the frequencies in these isotopes are predicted to vary with $\alpha$ according to $\Delta |\nu{\sss BA} |= (\pm 2 \times10^{15}\,\mathrm{Hz})\,\Delta\alpha/\alpha$, where the sign is positive for $^{162}$Dy and negative for $^{164}$Dy~\cite{Leefer.2013,Dzuba.2008,Flambaum.2009b}.


An experiment only limited by counting statistics would achieve a frequency uncertainty of $\approx 10$~mHz in one day of data taking. This correlates to a predicted sensitivity of  $|\dot{\alpha}/\alpha| \approx 5\times10^{-18}$~yr$^{-1}$ for two one-day measurements of the frequencies taken one year apart~\cite{Nguyen.2004}. The current measurement precision is limited, however, by systematic errors related to imperfections in the experimental environment. Because some of these systematics have the potential to mimic or mask a variation of $\alpha$, they must be properly accounted for in the final analysis. One of these systematics is the ac-Stark effect due to the electric-dipole interaction of either state $A$ or $B$ with additional states. An overview of other likely systematics can be found in Refs.~\cite{Nguyen.2004,Cingoz.2007,Leefer.2013}. 
 
\section{ac-Stark effect} \label{sec:ac_Stark_section}
\noindent  
The response of an atom to an external electric field is described by the linear electric polarizability. Specifically, the ac-Stark shift of an electronic energy level $|m\rangle$ due to an oscillating electric field with root-mean-square (rms) amplitude $E$ and frequency $\nu$ is given by
	\begin{equation}
	 \Delta W_m / h = -\frac{1}{2} \balpha_{m}(\nu) E^2,
	 \label{peq:1}
	\end{equation}
where $\balpha_{m}(\nu)$ is the electric-dipole polarizability\footnote{Note that there are different conventions used in the literature for the polarizability. In this work we use the rms-electric field to maintain consistency with the definition of dc polarizabilities.} of level $m$~\cite{Bonin1994}. Note that a bold $\balpha$ will refer to polarizabilities and should not be confused with the fundamental constant $\alpha$. The polarizability of an atomic state $|m\rangle$ can be written in the form~\cite{Geremia.2006,DounasFrazer.2010,Auzinsh.2010}
			\begin{equation}
			\begin{split}
			\label{peq:2}
			\balpha_{m}(\nu) = &\balpha^{(0)}_{m}(\nu) +  i \balpha^{(1)}_{m}(\nu)\frac{M}{F} (\bepsilon \times \bepsilon^*)\cdot \mathbf{\hat{z}}\,+ \\
		 							  & \balpha^{(2)}_{m}(\nu) \frac{3 M^2 - F (F+1)}{F(2F-1)}\frac{3|\bepsilon\cdot\mathbf{\hat{z}}|^2 - 1}{2},
		 	\end{split}
			\end{equation}
where $\bepsilon$ is the polarization vector and $\nu$ is the frequency the  of the electric field, $F$ is the total electronic angular momentum of state $|m\rangle$, $M$ is the projection of angular momentum along the chosen quantization axis, $\mathbf{\hat{z}}$, and the quantities $\balpha^{(0)}(\nu),\,\balpha^{(1)}(\nu),$ and $\balpha^{(2)}(\nu)$ are, respectively, the \textit{scalar}, \textit{vector}, and \textit{tensor} polarizabilities of state $|m\rangle$. These three quantities allow us to characterize ac-Stark shifts in a form independent of the experimental geometry or magnetic sublevel.

The differential ac-Stark shift of the $B \rightarrow A$ transition frequency is 
\begin{equation}
\begin{split}\label{peq:3}
	\Delta \nu_{\sss BA} &= \left(\Delta W_{\sss B} - \Delta W_{\sss A} \right) / h \\
											&= - \frac{1}{2}\left[\balpha_{\sss B}(\nu) - \balpha_{\sss A}(\nu)\right] E^2.
\end{split}
\end{equation}
The even-mass-number isotopes of Dy have zero nuclear spin, and their total angular momentum is $F = 10$ for both states $A$ and $B$. The odd-mass-number isotopes, $^{163}$Dy and $^{161}$Dy, have nuclear spin $I = 5/2$ with the corresponding range of total angular momenta $F = 15/2,17/2, \dots, 25/2$ for $A$ and $B$. In the case of a linearly polarized electric field, where $\bepsilon\times\bepsilon^*=0$ we can write 
for transitions with $M_A = M_B = M$:
		\begin{align}
			\label{peq:4}
			\Delta \nu_{\sss BA} =  -&\frac{1}{2} \left[\balpha^{(0)}_{\sss B}(\nu) - \balpha^{(0)}_{\sss A}(\nu)\right]E^2 - \\
			& \frac{1}{2}\left[\balpha^{(2)}_{\sss B}(\nu) - \balpha^{(2)}_{\sss A}(\nu)\right]				\frac{3 M^2 - F (F+1)}{F(2F-1)} E^2.	\nonumber
		\end{align}
%
We distinguish between two contributions to the differential polarizabilities in Eq.~\eqref{peq:4} based on their dependence in the frequency $\nu$. Radio-frequency (rf) spectroscopy of levels $A$ and $B$ is performed with an electric field that is nearly resonant with the $B\rightarrow A$ transition. We assume that the energy splitting between states $A$ and $B$ is much smaller than the splitting between $A$ or $B$ and any third state. With this we distinguish between the \textit{resonant} or \textit{two-state} electric-dipole interaction between states $A$ and $B$, and the \textit{off-resonant} electric-dipole interaction of states $A$ and $B$ with all other states. 

The resonant contribution to the differential polarizability is odd with respect to detuning from resonance, and for a perfectly resonant electric field it gives no contribution to ac-Stark shifts, neglecting the small Bloch-Siegert shift~\cite{Nguyen.2004}, which is related to the `counter-rotating' frequency component of the oscillating electric field~\cite{Budker.2008}. 

The off-resonant contribution to the differential polarizability, however, is approximately constant near these radio frequencies and can lead to systematic errors in the rf spectroscopy of transitions between $A$ and $B$. We make the distinction between the two contributions explicit by writing Eq.~\eqref{peq:4} as
	\begin{align}
	\label{peq:5}
		\Delta \nu_{\sss BA} = & \|d_{\sss BA}\|^2 \frac{|\langle F M 1 0|F M\rangle|^2}{2 F +1} f\left( \nu_{\sss BA}, \nu\right) E^2 - \\
	     								   & \frac{1}{2}\left(\overline{\balpha}^{(0)}_{\sss BA}(\nu) + \overline{\balpha}^{(2)}_{\sss BA}(\nu)\frac{3 M^2 - F (F+1)}{F(2F-1)}\right) E^2, \nonumber
	\end{align}
where the first term contains the resonant electric dipole interaction, and the second term contains the off-resonant interaction.
Here $\overline{\balpha}^{(0)}_{\sss BA}(\nu)$ and $\overline{\balpha}^{(2)}_{\sss BA}(\nu)$ are the differential scalar and tensor polarizabilities that include only the off-resonant contributions, $\|d_{\sss BA}\|$ is the reduced dipole matrix element between states $A$ and $B$, and 
\begin{equation}
\textstyle f\left( \nu_{\sss BA}, \nu\right) =  \text{Re}\left[ \frac{1}{\nu_{\sss BA} - \nu - i \frac{\Gamma_{\sss AB}}{2\cdot2\pi}} + \frac{1}{\nu_{\sss BA} + \nu + i \frac{\Gamma_{\sss AB}}{2\cdot2\pi}} \right]
\label{two_state_functional_dependence}
\end{equation}
contains the dependence of the resonant differential polarizability on the electric field frequency, $\nu$ \cite{BenjaminJ.Sussman.2011,DounasFrazer.2010,Budker.2008}. The quantity $\Gamma_{BA}/(2\pi)$ is determined by the radiative lifetimes of states $A$ and $B$ as well as transit effects; in our setup it is empirically determined to be $\approx45$ kHz.

From here on we will omit the frequency dependence of the off-resonant polarizabilities, whenever the electric-field frequency is comparable to $\nu_{\sss BA}$, i.e. $\overline{\balpha}^{(0,2)}_{\sss BA}(\nu \approx \nu_{\sss BA} ) = \overline{\balpha}^{(0,2)}_{\sss BA}$. To facilitate further discussion we introduce the compact notation
\begin{equation}
	\begin{split}
		d_{\ssBA}^2 &:= \|d_{\sss BA}\|^2 \frac{|\langle F M 1 0|F M\rangle|^2}{2 F +1}, \\
		\overline{\balpha}_{\ssBA} &:=\overline{\balpha}^{(0)}_{\sss BA} + \overline{\balpha}^{(2)}_{\sss BA}\frac{3 M^2 - F (F+1)}{F(2F-1)}.
	\end{split}
\label{eq:intruduce_short_notation}
\end{equation}
This allows us to rewrite Eq. \eqref{peq:5} as
	\begin{equation}
	\begin{split}
	\label{peq:5b}
		\Delta \nu_{\ssBA} = & d_{\ssBA}^2 f\left( \nu_{\sss BA}, \nu\right) E^2 - \frac{1}{2} \overline{\balpha}_{\ssBA} E^2,
	\end{split}
	\end{equation}
where the dependence on $M$ is implicit, $d_{\ssBA}$ is the dipole matrix element, and $\overline{\balpha}_{\ssBA}$ is the total differential polarizability for only the off-resonant contributions. 

A measurement of the reduced dipole matrix element is presented in Ref.~\cite{Budker.1994}. 
We use this value to calibrate the electric field amplitudes in our experiment. With these we determine the total off-resonant differential polarizabilities, $\overline{\balpha}_{\ssBA}$, from ac-Stark shifts of the $B \rightarrow A$ transition. The off-resonant scalar and tensor differential polarizabilities are determined by measuring $\overline{\balpha}_{\ssBA}$ for different Zeeman transitions.

\section{Measurements}
\noindent
\subsection{Suppressing magnetic-field instabilities} \label{sec:Suppressing_magnetic_field_instabilities}
\noindent 
%
%
To determine $\overline{\balpha}_{\sss BA}$ unambiguously we apply a magnetic field of sufficient strength to fully resolve the Zeeman structure of the transition. This enables us to measure Stark shifts of individual Zeeman transitions between states $A$ and $B$. The quantization axis is chosen to coincide with the linear polarizations of the applied electric field, such that only $M_{\sss B} = M_{\sss A}$, $\Delta M = 0$ transitions are observed. The change in the transition frequency due to the magnetic field is given by
\begin{align} 
 \nu_{\sss Z}(M) & = M \cdot \mu_B / h \cdot g_{{\scriptscriptstyle BA}}\cdot H \\& \approx M \cdot 220 \text{ [Hz/mG]} \cdot H[\textrm{mG}], 
\end{align}
where $\nu_{\sss Z}(M)$ is the differential Zeeman shift of the magnetic sublevel $M$ in the transition $B\rightarrow A$,  $\mu_B$ is the Bohr magneton, $g_{{\scriptscriptstyle BA}} = g_{\text{B}} - g_{\text{A}} = 0.157$ is the difference in g-factors for the states $A$ and $B$~\cite{kramida.2012}, and $H$ the strength of the applied magnetic field.

Over the course of a typical measurement ($\approx 20$  min) at constant temperature, the magnetic field drifts by $\approx 0.03$ mG; leading to drifts of  $\approx 70$ Hz for the measured transition frequency between the $M=10$ sublevels. As states $A$ and $B$ have the same total angular momentum $F$, the magnetic field insensitive $M_\text{B}=0 \rightarrow M_\text{A}=0$ transition is forbidden \cite{Budker.2008}. Instead of relying on the magnetic field insensitive transition, we suppress the magnetic field related uncertainties by measuring the transition frequencies of the $+M$ and $-M$ sublevels nearly simultaneously.

The frequency $\nu_{\scriptscriptstyle BA}(M)$ for the  $B \rightarrow A$  transition between sublevels $M$ under the influence of a magnetic and an electric field is given by:
\begin{equation}
	\begin{split}
		\nu_{\ssBA}(M)  = & \nu_{\ssBA} + \nu_{\sss Z}(M) - \frac{1}{2}  \overline{\balpha}_{\sss AB} E^2 +\\
		&   d^2_{\scriptscriptstyle BA} f\left[ \nu_{\sss BA}+\nu_{\sss Z}(M), \nu\right] E^2\text{.} 
	\end{split}
\end{equation}
Note that under our assumption that the separation between states $A$ and $B$ is much smaller than their separation from any other states $\overline{\balpha}_{\sss AB}$ is also insensitive to magnetic fields. 

The Zeeman shift, $\nu_{\sss Z}(M)$, changes sign with respect to the sign of $M$, while the Stark shift does not, thus our measured quantity is the magnetic field insensitive average of the $\pm M$ Zeeman transitions: 
\begin{align}
	\bar{\nu}_{\ssBA} & =  \frac{1}{2} \left[ \nu_{\sss BA}(+M) + \nu_{\sss BA}(-M) \right] \nonumber \\
		 			& =  \nu_{\ssBA}  - \frac{1}{2}  \overline{\balpha}_{\sss BA} E^2 + \label{eq:ac-stak_shift_zeeman_cancels_full} 		 \\
		 			& \phantom{=} \frac{1}{2} d^2_{\sss BA} \big\{ f\left[ \nu_{\sss BA}-\nu_{\sss Z}(M), \nu \right] + f\left[ \nu_{\sss BA}+\nu_{\sss Z}(M), \nu \right]  \big\} E^2 . \nonumber
\end{align} %
Two effects may make the cancellation of magnetic fields imperfect. The $-M$ and $+M$ transitions are measured sequentially, $\approx 1$~s apart from each other. In a magnetic field drifting at a rate $\dot{H}$~[mG/s]  the average $\pm M$ frequency will have a systematic shift of $\approx (1/2)\cdot 220\cdot M \dot{H} \Delta t$~[Hz], where $\Delta t$ is the measurement interval.

Second, the Zeeman shift in the two-state term cancels incompletely due to the presence of $\nu_{\sss Z}$ in the argument of $f\left(\nu_{\sss BA}\pm \nu_{\sss Z}, \nu\right)$, Eq.~\eqref{two_state_functional_dependence}. This imperfect cancellation is in practice negligible as long as the difference between the Zeeman shifted transition frequencies,  $\nu_{\ssBA}(\pm M)$, and the electric field frequency $\nu$ is much larger in magnitude than the change in the Zeeman shift $\delta \nu_{\sss Z}(M)$ due to varying magnetic fields over the course of the measurement [typically $|\nu_{\ssBA}(\pm M) - \nu|  \geq 10^{6}$ Hz vs $|\delta \nu_{\sss Z}(M)| < 4\times10^{3}$ Hz].

\begin{figure}[t]
\center
 \includegraphics[width=1\columnwidth]{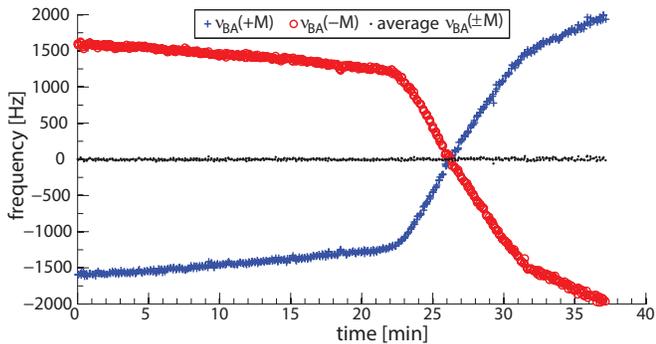}
\caption{Frequency of the $M=+10$, $M=-10$ and the $M=\pm 10$ average Zeeman transition, offset for presentation. Over the course of this measurement the individual Zeeman transition frequencies changed by $\approx 3500$~Hz while the $\pm M$ average changed by $\approx8$~Hz, demonstrating a suppression of $\approx 450$. The actual separation between the $M=+10$ and $M=-10$ transition is $\approx3 \times 10^6$ Hz. These data were taken while the interaction region was cooled toward liquid nitrogen temperatures. The relatively large drift in magnetic field of ${>1}$~mG is believed to be due to thermoelectric currents induced by temperature gradients between dissimilar metals.}
\label{fig:zeemansuppression}
\end{figure}

We find empirically that drifts of the Zeeman shifts between the measurements for $+M$ and $-M$  are suppressed by up to a factor of $10^3$. An example of the cancellation can be seen in Fig.~\ref{fig:zeemansuppression}.
\subsection{Frequency Measurement}\label{sec:Procedure}
\begin{figure}[t]
	\centerline{ \includegraphics[width=1\columnwidth]{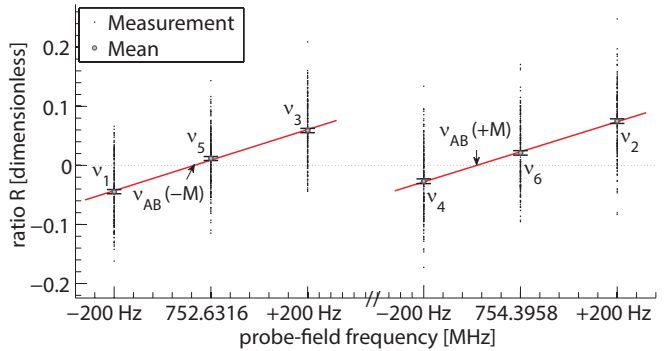}}
	\caption{The Zeeman structure of the $B \rightarrow A$ transition is resolved with a 550-mG field. The ratio R, as determined by Eq.~\eqref{eq:define_ratio_behavior}, is measured at the probe field frequencies $\nu_i$ for $i = 1,....,6$.}
	\label{fig:measure_scheme}
	\end{figure}

\noindent 
The transition frequencies are measured by scanning a frequency-modulated probe field near resonance. We perform lock-in detection of the 564-nm fluorescence light emitted by Dy atoms during the final decay step of level $A$ to the ground state~\cite{Cingoz.2005} (Fig.~\ref{fig:popscheme}). Near resonance, the lock-in signals at the first and second harmonic of the 10-kHz modulation frequency are approximated by a linear function that crosses zero on resonance, and a constant function, respectively. The quantity from which we determine the resonance frequency is the ratio of these signals, as it is insensitive to changes in signal size that may arise from fluctuations in the density of excited-state atoms \cite{Cingoz.2005}. For small detunings, this ratio is given by 
\begin{equation}
 R(\nu) = S \left(\nu - \nu{\ssBA} \right),
\label{eq:define_ratio_behavior}
\end{equation}
where $\nu$ is the probe-field frequency and $S$ is the empirically determined slope (Fig \ref{fig:measure_scheme}).

We measure $R$ repeatedly for three probe-field frequencies at intervals of 200 Hz near the expected resonance frequency for each Zeeman transition (Fig. \ref{fig:measure_scheme}). The slope $S$ is determined from the linear least-squares fit to the mean signal ratios. With the slope $S$, each measurement of $R$ is converted into a transition frequency via the relation
	\begin{equation}
	  \nu_{\sss BA} = \nu - \frac{R(\nu)}{S}.
	\label{eq:slope_and_intercept}
	\end{equation}

\section{Experimental Setup} 
\begin{figure*}[t]
		\centerline{ \includegraphics[width=1.0\textwidth]{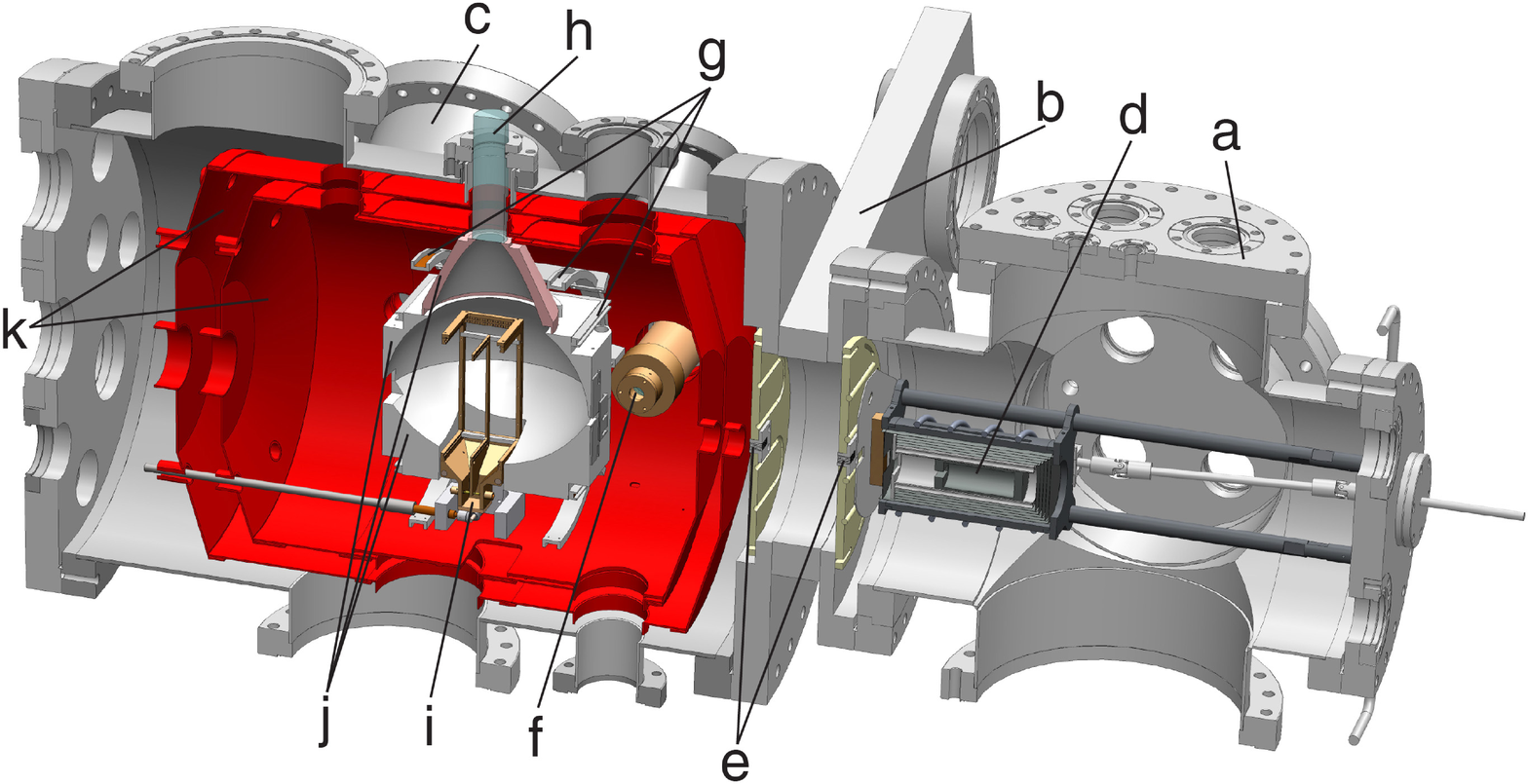}}
		\caption{Section view of the atomic beam apparatus.\\
		a) oven chamber;
		b) gate valve;
		c) interaction-region chamber;
		d) Dy oven;
		e) vacuum chokes;
		f) laser access/in-vacuum polarizer;
		g) magnetic coils;
		h) light pipe;
		i) rf electrodes;
		j) light-collection mirrors;
		k) two-layer magnetic shielding}
		\label{fig:apparatus}
\end{figure*}
\noindent
The rf spectroscopy is performed with an atomic-beam apparatus as depicted in Fig.~\ref{fig:apparatus}.
A thermal beam of Dy atoms is emitted by an effusive oven, operated at $\approx$1400 K. Two vacuum chokes are used to facilitate a pressure differential from $10^{-7}$ to $10^{-9}$ Torr between the oven and interaction regions. Excitation of atomic states is performed with 833-nm and 669-nm light with an adiabatic-passage population technique that uses diverging laser beams matched to the transverse divergence of the atomic-beam~\cite{Nguyen.2000}. The 669-nm light is generated by a Coherent CR-699 dye laser using dicyanomethylene (DCM) dye and pumped by a Coherent Innova-300 argon-ion laser. Two sources have been used to generate the 833-nm light. One source is a custom built master-oscillator power-amplifier system, consisting of a Littrow-configuration extended-cavity diode laser (ECDL) whose output is amplified with a tapered amplifier (TA). The other source is a Coherent CR-899 Ti:Sapphire laser pumped by a Coherent Innova-400 argon-ion laser. Typical powers incident at the atoms are 150 mW (669 nm) and 250 mW (833 nm).

The laser and rf interaction regions are surrounded by two layers of magnetic shielding that suppress external fields to below 0.5~mG. Three pairs of coils within the shielded volume provide additional control over magnetic fields in all directions. 

\begin{figure}[b]
\centerline{ \includegraphics[width=1\columnwidth]{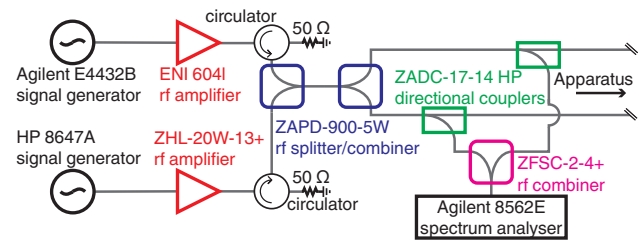}}
\caption{Radio-frequency setup for the measurement of ac-Stark shifts. The probe field and a second off-resonant electric field are provided by two separately amplified signal generators. Radio-frequency circulators isolate the amplifiers from reflections from the rf region. Rf splitter and combiner ensure a symmetric feed of the signals into the rf region. The directional couplers allow us to monitor the rf power before the apparatus.
}
\label{fig:rf_setup}
\end{figure}

Two rf-electric fields 
are provided by two signal generators with separate amplifiers (Fig. \ref{fig:rf_setup}). The amplified rf fields are combined and fed to the interaction region, incorporating a rectangular electrode that is surrounded on all sides by a grounded box. This nested rectangular configuration supports a transverse electromagnetic mode and provides a homogeneous field up to 1 GHz. Symmetric feeding ensures a standing wave configuration of the field.  The `surfaces' of the electrodes and box are defined by a series of parallel gold-plated 0.002-in Be-Cu wires stretched across gold-plated copper frames at 2-mm intervals. This electrode design is effectively transparent to atoms and photons, while effectively solid for the wavelength range we operate in ($> 30$~cm). A partial view of the electrode frames, without wires, can be seen in Fig.~\ref{fig:apparatus}.

Atoms that are excited from $B$ to $A$ decay back to the ground state, with a lifetime of $\approx8$ \textmu s, via several channels. In one of these channels the final step includes the emission of a 564-nm photon. Three concave mirrors made from polished aluminum focus these 564-nm photons into a Pyrex lightpipe, which guides the light into a photomultiplier tube (PMT). The PMT signal is processed with a digital lock-in amplifier.
\begin{figure*}[t!] 
 \includegraphics[width=.67\textwidth]{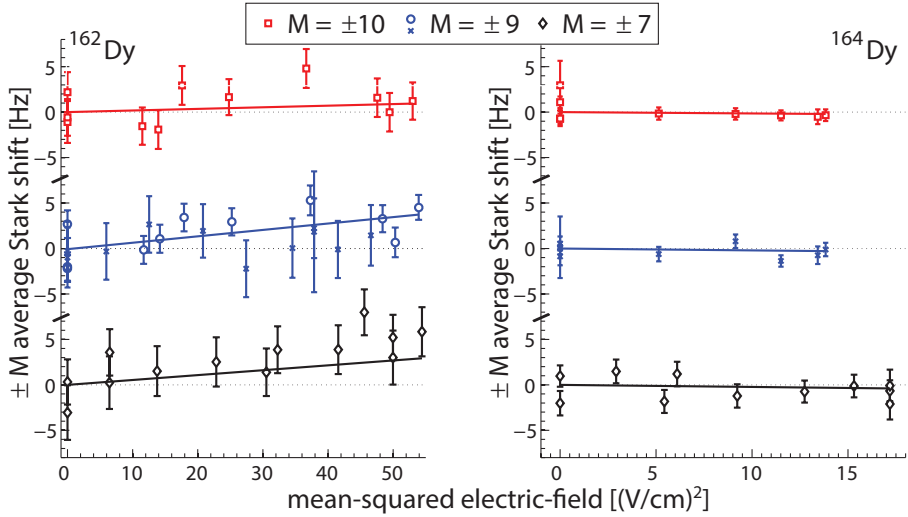}
\caption{The mean ac-Stark shift of the $\pm M$ transitions in $^{162}$Dy and $^{164}$Dy, as a function of mean-squared electric-field amplitude $E^2$. Lines of best fit are obtained from least-squares fit of Eq.~\eqref{eq:fit_function} to the data. Measurements for different Zeeman transitions are offset from each other for display purposes. Note the difference in horizontal scale.
}
\label{fig:Vary_E}
\end{figure*}
The magnetic field coils, rf electrodes, and light-collection mirrors constitute a single assembly. This assembly is clamped to two copper rods inside the vacuum chamber. The copper rods pass to the exterior of the vacuum setup, thermally and electrically isolated from the chamber by ceramic feedthroughs. These copper `cold fingers' allow the rf electrodes and surrounding light-collection mirrors to be heated or cooled without introducing heating elements or cryogens into the vacuum environment. The temperatures of the light-collection assembly, oven, and vacuum chamber are continuously monitored with thermocouples.
\section{Results - ac-Stark shift}\label{sec:ac-results}
\begin{figure}[b]
\includegraphics[width=1\columnwidth]{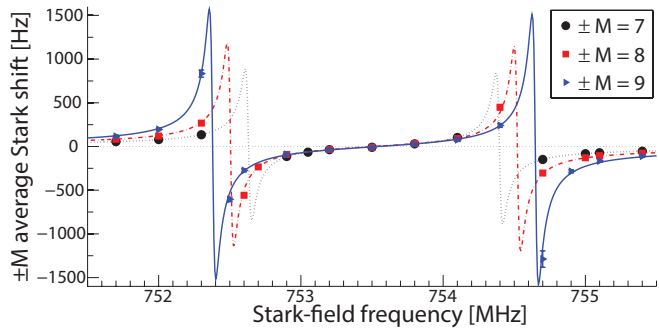}
\caption{Average ac-Stark shift of the $\pm M$ sublevels, as a function of the Stark-field frequency $\nu_S$. The zero crossings at the centers of the dispersive resonances indicate the approximate location of the individual Zeeman transition frequencies. The mean-squared amplitude of the Stark field is $\approx9$~(V/cm)$^2$.}
\label{fig:Vary_nu}
\end{figure}

\noindent
We determine the differential off-resonant polarizability,
 $\overline{\balpha}_{\sss AB}$, from the ac-Stark shift of the $\pm M$ average transition frequency, $\Delta\bar{\nu}_\text{BA}$. The transition is ac-Stark shifted by a second, unmodulated, oscillatory electric field with frequency $\nu_S$ and mean-squared amplitude $E^2$. This field is referred to as the `Stark field'. We measure the Stark shifts as both a function of Stark-field amplitude (Fig.~\ref{fig:Vary_E}) and Stark-field frequency (Fig.~\ref{fig:Vary_nu}).
The shift $\Delta \bar{\nu}_{\sss BA}$ is given by 
	\begin{equation}
	\begin{split}
					\Delta\bar{\nu}_{\sss BA} =  &\left(\frac{d_{\scriptscriptstyle BA}^2}{2} \right. \big\{ f\left[ \nu_{\sss BA}-\nu_{\sss Z}(M), \nu_S \right] + \\ &\left. f\left[ \nu_{\sss BA}+\nu_{\sss Z}(M), \nu_S \right] \big\} - \frac{\overline{\balpha}_{\sss AB}}{2} \right)E^2.
			 \label{eq:fit_function} 
	\end{split}
	\end{equation}
For the measurements as a function of Stark-field    amplitude we choose $\nu_S$ so that the two-state shift cancels in the average shift of the $\pm M$ sublevels (neglecting the Bloch-Siegert shift discussed below). This occurs at the frequency $\nu_S = \bar{\nu}_{\scriptscriptstyle BA}|_{E=0}$ (around $\approx 753.5$ MHz for $^{164}$Dy, see Fig.~\ref{fig:Vary_nu}). The remaining shift is only given by the differential off-resonant polarizability and the Bloch-Siegert shift, which is on the order of ${\scriptstyle\lesssim}\,30$ mHz and accounted for in the analysis. These data and the resulting least-squares fits to Eq.~\eqref{eq:fit_function} are shown in Fig.~\ref{fig:Vary_E}.

To determine $\overline{\balpha}_{\sss AB}$ precisely, the frequency and amplitude of the Stark-field inside the chamber have to be controlled.
%
The signal generators for the Stark- and probe-fields are referenced to a Cs frequency standard, which ensures frequency uncertainty better than $2$ Hz and reduces the systematic uncertainty in the total differential polarizability to $< 0.2 \text{ mHz}/\left(\mathrm{V}/\mathrm{cm}\right)^2$. This is negligible compared to the statistical uncertainty, which is on the order of $100 \text{ mHz}/\left(\mathrm{V}/\mathrm{cm}\right)^2$.

The mean-squared amplitude of both fields is monitored outside the apparatus by a spectrum analyzer (Fig~\ref{fig:rf_setup}). It is calibrated to the field inside the apparatus via the two-state shift of the $B\rightarrow A$ transition. This shift is given by the difference of the $\pm M$ transition frequencies
\begin{align}\label{eq:ecalibration}
&\Delta\big[\nu_{\sss BA}(+M) - \nu_{\sss BA}(-M)\big] = \\ &d_{\sss BA}^2 \big\{ f\left[ \nu_{\sss BA}+\nu_{\sss Z}(M), \nu_S \right] -   f\left[ \nu_{\sss BA}-\nu_{\sss Z}(M), \nu_S \right] \big\} E^2. \nonumber
\end{align}
We calculate $d_{\ssBA}^2$ from ${\|d_{\sss BA}\|^2 = 19.2^2\ \mathrm{ kHz}^2/\left(\mathrm{V}/\mathrm{cm}\right)^2}$ \cite{Budker.1994}.
These calibrations show that the maximum mean-square amplitudes of the Stark-field inside the apparatus are significantly different between isotopes for comparable maximum set amplitudes: 
15 (V/cm)$^2$ for $^{164}$Dy at 753.5 MHz, and 50 (V/cm)$^2$ for $^{162}$Dy at 234.7 MHz, Fig~\ref{fig:Vary_E}. This large difference in maximum value is hypothesized to be due to a substantial frequency dependence of the impedance mismatch between the rf transmission line and the electric-field plates. 

For the measurements as a function of Stark-field frequency, we fit Eq.~\eqref{eq:fit_function} to a set of $\nu_S$, $\Delta\bar{\nu}_{\sss BA}$ pairs, where $\overline{\balpha}_{\sss AB}$, and $E^2$ are free parameters of the fit.

The total differential off-resonant polarizabilities are determined for the $M = 7,\,8,\,9,\,\mathrm{and}\,10$ transitions in $^{164}$Dy and $^{162}$Dy, as shown in Fig.~\ref{fig:Vary_E} and Fig.~\ref{fig:Vary_nu}, and fit via least-squares according to the relationship
\begin{equation}\label{eq:pol_decomp}
		\overline{\balpha}_{\sss BA} = \overline{\balpha}_{\sss BA}^{(0)} + \overline{\balpha}^{(2)}_{\sss BA} \frac{3 M^2 - F (F+1)}{F(2F-1)}.
\end{equation}
The differential scalar and tensor off-resonant polarizabilities are found to be
\begin{align}
\overline{\balpha}_{\ssBA}^{(0)} &= \left \{ \begin{array}{cc}
  \phantom{-}220\ (83)_\text{stat}\, (8)_\text{sys}\,\,\text{mHz}/\left( \mathrm{V}/\mathrm{cm}\right)^2 & ^{162}\mathrm{Dy}\\
  \phantom{-2} 90\ (60)_\text{stat}\, (7)_\text{sys}\,\, \text{mHz}/\left( \mathrm{V}/\mathrm{cm}\right)^2 & ^{164}\mathrm{Dy}\\
	\phantom{-}180\ (45)_\text{stat}\, (8)_\text{sys}\,\,\text{mHz}/\left( \mathrm{V}/\mathrm{cm}\right)^2 & \mathrm{combined},\\
\end{array}\right. \nonumber\\ \label{eq:result_scalar_and_tensor_pol} \\
\overline{\balpha}_{\ssBA}^{(2)} &= \left \{ \begin{array}{cc}
- 235\ (123)_\text{stat}\, (6)_\text{sys}\,\,\text{mHz}/\left( \mathrm{V}/\mathrm{cm}\right)^2 & ^{162}\mathrm{Dy}\\
 \phantom{2} - 7\ \phantom{1}(82)_\text{stat}\, (3)_\text{sys}\,\,   \text{mHz}/\left( \mathrm{V}/\mathrm{cm}\right)^2 & ^{164}\mathrm{Dy}\\
	-163\ \phantom{1}(65)_\text{stat}\, (5)_\text{sys}\,\,\text{mHz}/\left( \mathrm{V}/\mathrm{cm}\right)^2 & \mathrm{combined}.\\
\end{array}\right.  \nonumber
\end{align} 
In order to obtain the combined results we fit Eq.~\eqref{eq:pol_decomp} to the differential off-resonant polarizabilities of $^{162}$~Dy and $^{164}$~Dy together. We do this with the assumption that the off-resonant polarizabilities are independent of the isotope.

Assuming that the differential polarizability arises due to the electric-dipole interaction of state $A$ or $B$ with only one other state, the combined ratio $\overline{\balpha}_{\ssBA}^{(2)}/\overline{\balpha}_{\ssBA}^{(0)} = -0.9(4)$ is consistent with a total angular momentum of $J = 9$ or $J = 11$ for the partner state~\cite{Auzinsh.2010}. The closest documented state is the odd-parity, $4f^95d^26s\,J = 9$ state at 19,558 cm$^{-1}$~\cite{kramida.2012}. The energy of this state relative to the energy of state $A$ (19,797 cm$^{-1}$) is consistent with the sign of the scalar polarizability. The magnitude of the scalar polarizability would require a reduced dipole-matrix element between this state and state $A$ of $\|d\| = 6.9(0.9)$~ea$_0$. 

We also cannot rule out the existence of close lying states not listed in the spectroscopic databases. Studying Raman transitions within the Zeeman manifold of either state $A$ or $B$ would allow their individual contributions to the differential polarizability to be determined and will provide more information about the off-resonant states.

If the off-resonant differential polarizabilities are due to the interaction with states that are far removed from state A and B, the polarizabilities presented in Eq.~\eqref{eq:result_scalar_and_tensor_pol} are effectively dc-polarizabilities. In this case the off-resonant polarizabilities 
 add a significant contribution to the total dc polarizabilities of states $A$ and $B$ in $^{162}$Dy and $^{164}$Dy. Using the value for $\|d_{\sss BA}\|$ from Ref.~\cite{Budker.1994} the two-state dc differential polarizabilities are given by~\cite{Bonin1994}:
\begin{align}
{\balpha}_{\ssBA}^{(0)}( \text{dc} ) - \overline{\balpha}_{\ssBA}^{(0)}( \text{dc} ) = \left \{ \begin{array}{cc}
100\ (14)\,\,\text{mHz}/\left( \mathrm{V}/\mathrm{cm}\right)^2  & ^{162}\mathrm{Dy}\\
\phantom{1}31\ \phantom{1}(4)\,\,\text{mHz}/\left( \mathrm{V}/\mathrm{cm}\right)^2  & ^{164}\mathrm{Dy}\\
\end{array}\right. ,\nonumber \\ \label{eq:two_state_DC_pols}\\
{\balpha}_{\ssBA}^{(2)}( \text{dc} )  - \overline{\balpha}_{\ssBA}^{(2)}( \text{dc} )  = \left \{ \begin{array}{cc}
172\ (24)\,\,\text{mHz}/\left( \mathrm{V}/\mathrm{cm}\right)^2  & ^{162}\mathrm{Dy}\\
\phantom{1}54\ \phantom{1} (8)\,\,\text{mHz}/\left( \mathrm{V}/\mathrm{cm}\right)^2  & ^{164}\mathrm{Dy}\\
\end{array}\right. .\nonumber 
\end{align}
These values are comparable in size to the off-resonant contributions in \eqref{eq:result_scalar_and_tensor_pol}. Note though that even if the off-resonant polarizabilities are given by the interaction with far removed states, they do not add significantly to the DC polarizabilities for all transitions B$\rightarrow$A for all isotopes. For transitions where the separation is smaller, like the 3.1~MHz transition in $^{163}$Dy \cite{Budker.1994}, the two-state DC polarizabilities are larger due to the reduced separation, and the off-resonant polarizabilities only contribute negligibly to the total DC polarizability. 


%
\section{Blackbody radiation} \label{sec:BBR_part}
\noindent
The electric field of blackbody radiation (BBR) from the environment can cause ac-Stark shifts of atomic energy levels~\cite{Levi.2004}. The mean-squared electric-field amplitude of blackbody radiation in a frequency interval $d\nu$ in a vacuum surrounded by an enclosure of uniform temperature $T$ is given by Planck's law as~\cite{Itano.1982}
	\begin{equation}
		E_{\scriptscriptstyle \text{BBR}}^2 (T,\nu) d \nu = 32 \pi^2 \frac{ h \nu^3}{c^3} \frac{300^2}{e^{\frac{h\nu}{kT}}-1} d\nu\quad  \left( \mathrm{V}/\mathrm{cm}\right)^2 ,
		\label{eq:BBR_T_and_freq_dep}
	\end{equation}
where $k$ is the Boltzmann constant, $c$ is the speed of light, $T$ is the absolute temperature in Kelvin, and all the quantities on the right-hand side are assumed to be in the  centimeter-gram-second (cgs) units; the factor of $300^2$ converts the expression to units of  (V/cm)$^2$ .

Blackbody radiation is isotropic with no preferred axis. The ac-Stark shift of the $B \rightarrow A$ transition is therefore given by the differential scalar polarizability and the integral over all frequencies $\nu$,
		\begin{align}
			\Delta \nu_{{\scriptscriptstyle BA}}(T) = 
			 - \frac{1}{2}\int_0^{\infty}  E^2_{\scriptscriptstyle \text{BBR}}(T,\nu) \left[\balpha_{\sss B}^{(0)}(\nu) - \balpha_{\sss A}^{(0)}(\nu)\right] d \nu.
			 \label{eq:BBR_scalar}
		\end{align} 
We again differentiate between the BBR related shift due to the interaction between $A$ and $B$ and the contribution from off-resonant states by writing
		\begin{align}\label{eq:BBR_scalar2} \textstyle
			\Delta \nu_{{\scriptscriptstyle BA}}(T) = 
			 \int_0^{\infty}  E^2_{\scriptscriptstyle \text{BBR}}(T,\nu) \left[ d^2_{\ssBA} f\left(\nu_{\sss BA},\nu \right) -\frac{1}{2} \overline{\balpha}_{\sss BA}^{(0)}(\nu)\right] d \nu.
		\end{align} 
The separation between states $A$ and $B$ is small compared to characteristic frequencies of BBR at 300 K. From the value for the dipole matrix element presented in Ref.~\cite{Budker.1994}, we can calculate the BBR radiation shift due to $d^2_{\ssBA}$ around 300 K to be
		\begin{align} 
		{ \frac{d}{dT} }\int_0^{\infty}  E^2_{\sss \text{BBR}}(T,\nu) d^2_{\ssBA}   f\left( \nu_{\sss BA}, \nu\right) d \nu |_{T=300\text{K}}  <10^{-10}{\textstyle  \frac{\text{Hz}}{{K}}}, 
		\end{align} 
which is negligibly small. 

It is important to stress that $\overline{\balpha}_{\ssBA}^{(0)}(\nu)$ in Eq.~\eqref{eq:BBR_scalar2} is a frequency dependent value, which we measured only for $\nu\approx 234.7$~MHz and $\nu\approx 753.5$~MHz. 
The dense level structure of Dy makes it probable that energy levels with strong electric-dipole coupling to state $A$ or $B$ exist within the thermal radiation spectrum, leading to unknown behavior of $\overline{\balpha}_{\ssBA}^{(0)}(\nu)$ in the range of BBR frequencies ($\nu \geq$~THz).

To determine $\overline{\balpha}_{\ssBA}^{(0)}(\nu)$ for BBR, we measure the transition frequency, $\nu_{\ssBA}$, as a function of temperature of the interaction region, which consists of the light-collection mirrors and electric-field plates as shown in Fig. \ref{fig:apparatus}.

The functional dependence of BBR induced shifts on temperature is generally unknown due to the temperature dependence of the BBR spectrum and the frequency dependence of $\overline{\balpha}_{\ssBA}^{(0)}$. A starting assumption, however, is that the energy splitting between $A$ or $B$ and other states is much larger than the characteristic energy of the BBR spectrum. In this approximation, the off-resonant scalar polarizability in Eq.~\eqref{eq:BBR_scalar2} is the same as that measured in Sec.~\ref{sec:ac-results} and we can write
\begin{equation}
 \Delta \nu_{\ssBA} = - \frac{1}{2} \overline{\balpha}_{\ssBA}^{(0)} \left(\frac{T}{300 \text{ K}} \right)^4 8.32^2\, \left( \mathrm{V}/\mathrm{cm}\right)^2.
\label{eq:BBR_parametrization}
\end{equation}
Here 8.32 (V/cm)$^2$ is the rms electric-field value of room temperature BBR.

\section{Results - Blackbody radiation shifts} \label{sec:results_BBR_shifts}
\begin{figure}[t]
		\centerline{ \includegraphics[width=1\columnwidth]{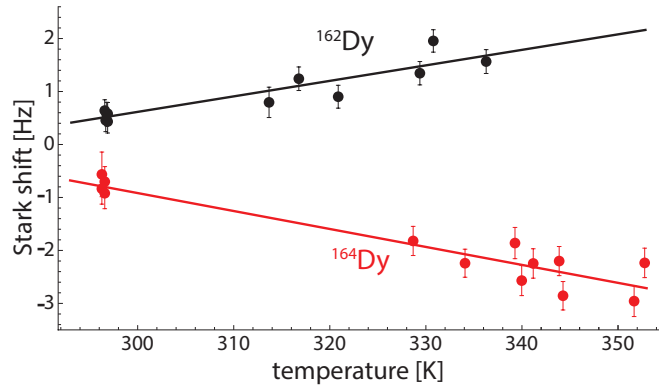}}
		\caption{Change in the frequency of the unresolved $B \rightarrow A$ transition as a function of interaction region temperature.}
		\label{fig:BBR_shifts}
\end{figure}

\noindent 
The transition frequency $\nu_{\scriptscriptstyle BA}$ for $^{164}$Dy and $^{162}$Dy was measured at zero magnetic field, with unresolved Zeeman structure, for temperatures of the interaction region ranging between 298~K and 352~K. 
Results are shown in Fig.~\ref{fig:BBR_shifts}, and the measured slopes are:
\begin{equation}
		\left.\frac{d}{dT} \nu_{{\scriptscriptstyle BA}}\right|_{300 \text{K}}= \left\{\begin{array}{cr}
		- 34(4) \text{ mHz/K} & \,^{164}\text{Dy}\phantom{.}\\
		+ 29(4) \text{ mHz/K} & \,^{162}\text{Dy}.
		\end{array}\right.
		\label{eq:BBR_shift_300K_164Dy}
\end{equation}
The signs of the measured frequency shifts are expected to be opposite for these two isotopes due to the different sign of the energy splitting between $A$ and $B$. The temperature range is too small to verify a $T^4$ dependence. Performing a linear expansion of Eq.~\eqref{eq:BBR_parametrization} around 300 K we find that the linear slopes correspond to differential scalar polarizabilities:
\begin{equation}
	\overline{\balpha}_{\sss BA}^{(0)} = \left\{\begin{array}{cr}
	 74\, (9)\  \text{mHz}/\left( \mathrm{V}/\mathrm{cm}\right)^2    & \,^{164}\text{Dy}\phantom{.}\\
	 63\, (9)\  \text{mHz}/\left( \mathrm{V}/\mathrm{cm}\right)^2    & \,^{162}\text{Dy}.
	\end{array}\right. 
	\label{eq:results_BBR_polarizability}
\end{equation}
These values are on the same order as the polarizabilities measured at radio frequencies, Eq.~\eqref{eq:result_scalar_and_tensor_pol}. However, these are not expected to be the same due to the possibility of many more atomic states contributing significantly to BBR shifts. For instance, in Dy the closest neighbor state at 19,558~cm$^{-1}$ is only $7$~THz removed from $A$ or $B$, while the spectrum of BBR at 300~K peaks at $24$~THz, with a full-width at half-maximum bandwidth of 27~THz. 
\section{Conclusion} \label{sec:implications}
\noindent
We have presented measurements of the off-resonant differential polarizabilities of states $A$ and $B$ in $^{162}$Dy and $^{164}$Dy for radio-frequency electric fields, and the effect of temperature on the transition frequency between these states in both isotopes.

Non-zero off-resonant polarizabilities could result in systematic errors in the measured value of $\nu_{\ssBA}$. In the experiment dedicated to constraining variation of $\alpha$ we are not concerned with the overall systematic error, but the \textit{stability} of the systematic error over the course of the experiment's lifetime. The stability of each ac-Stark related systematic is discussed in order to project the systematic limits on a search for variation of $\alpha$.

The probe electric field in the rf spectroscopy of the $B \rightarrow A$ transition has a typical mean-squared amplitude of $E^2 = 4.5$~(V/cm)$^2$. Assuming an instability of $\delta E^2 = 0.45$~(V/cm)$^2$, the two-state ac-Stark shift for a resonant probe field contributes a systematic frequency uncertainty of

\begin{align}
\sigma_{\nu_{\sss BA} }&= \| d_{\ssBA}\|^2 F(\nu_{\sss BA},\nu_{\sss BA})\delta E^2 \sum_M a_{\sss M}\frac{|\langle FM10|FM\rangle|^2}{2F+1}\notag\\
&= \left \{ \begin{array}{cc}
8\,\,\mathrm{mHz} & ^{162}\mathrm{Dy}\\
2\,\,\mathrm{mHz} & ^{164}\mathrm{Dy}\\
\end{array}\right.,
\end{align}
where the sum is over the normalized signal amplitudes, $a_{\sss M}$, of the unresolved Zeeman transitions. These amplitudes are measured with the Zeeman structure of the transition fully resolved.

The systematic uncertainty arising from off-resonant ac-Stark shifts is evaluated with the combined maximum-likelihood differential polarizabilities from Eq.~\eqref{eq:result_scalar_and_tensor_pol}:

\begin{equation}
\begin{split}
\sigma_{\nu_{\sss BA} }&= \frac{1}{2} \delta E^2 \sum_M a_{\sss M}\left(\overline{\balpha}_{\ssBA}^{(0)} + \overline{\balpha}_{\ssBA}^{(2)}\frac{3M^2 - F(F+1)}{F(2F-1)}\right)\\
&= 33\,\,\mathrm{mHz}.
\end{split}
\end{equation}
%
The temperature of the light collection mirrors was found to vary between 294~K and 298~K for measurements spanning over two years. The result of Eq.~\eqref{eq:BBR_shift_300K_164Dy} gives an estimate of the instability of $\nu_{\scriptscriptstyle BA}$ due to temperature dependent effects of
\begin{equation}
	\sigma_{\nu_{\scriptscriptstyle BA}} = 64\,\,\text{mHz}.
\end{equation}
Dysprosium atoms are also subject to the thermal radiation from the atomic-beam oven. The higher temperature of the oven (1400 K vs. 300 K) makes its radiation $\approx470$ times more intense. Due to the distance between the interaction-region and the oven, however, the intensity of oven BBR at the rf region is reduced by a factor $\approx10^{-4}$. The typical temperature variability of the oven is $\pm10$ K, and, neglecting the change in the frequency spectrum of BBR, the systematic uncertainty due to oven BBR radiation is 
\begin{equation}
	\sigma_{\nu_{\scriptscriptstyle BA}} = 16\,\,\text{mHz}.
\end{equation}
The four systematic uncertainties discussed are added in quadrature to evaluate the total ac-Stark related systematic uncertainty:
\begin{equation}
\sigma_{\nu_{\ssBA}} = 74\,\,\mathrm{mHz}.
\end{equation}
For two measurements of the transition frequencies $\nu_{\ssBA}$ one year apart this systematic uncertainty will not limit the experimental sensitivity to variation of $\alpha$ in each isotope down to the level of $|\dot{\alpha}/\alpha| = 2.6\times10^{-17}$~yr$^{-1}$, 
which is comparable to the present best limit~\cite{Rosenband.2008}. The ac-Stark systematic limit is currently dominated by the temperature instability of the interaction region. Improving the temperature is thus the best first option to reduce this systematic. Further studies of the BBR induced shifts may allow frequency measurements to be corrected for drifts of the interaction region temperature. Improving the stability of the rf-field amplitude will provide another increase in sensitivity. 

We have also shown that the off-resonant polarizabilities might contribute substantially to the differential dc-polarizability of levels $A$ and $B$ in $^{162}$Dy, and $^{164}$Dy.
In addition to helping constrain systematic uncertainties for searches of variation of $\alpha$ and Lorentz violation, this work provides spectroscopic information about the states $A$ and $B$ that can be used to test the atomic-structure calculations that are used in conjunction with these experiments~\cite{Dzuba.2008}.

\section*{Acknowledgments}
\noindent
{We thank Arman Cing\"{o}z, Valeriy Yashchuck, Alain Lapierre, and Tuan Nguyen for designing the Dy-atomic beam apparatus; Justin Torgerson and Ed Marti for the assistance with the MOPA system.  This project has been founded in part by NSF grant PHY-1068875, NIST, LANL, and FQXi.}



\bibliography{bibliography,bibliography2}

\begin{thebibliography}{35}%
\makeatletter
\providecommand \@ifxundefined [1]{%
 \@ifx{#1\undefined}
}%
\providecommand \@ifnum [1]{%
 \ifnum #1\expandafter \@firstoftwo
 \else \expandafter \@secondoftwo
 \fi
}%
\providecommand \@ifx [1]{%
 \ifx #1\expandafter \@firstoftwo
 \else \expandafter \@secondoftwo
 \fi
}%
\providecommand \natexlab [1]{#1}%
\providecommand \enquote  [1]{``#1''}%
\providecommand \bibnamefont  [1]{#1}%
\providecommand \bibfnamefont [1]{#1}%
\providecommand \citenamefont [1]{#1}%
\providecommand \href@noop [0]{\@secondoftwo}%
\providecommand \href [0]{\begingroup \@sanitize@url \@href}%
\providecommand \@href[1]{\@@startlink{#1}\@@href}%
\providecommand \@@href[1]{\endgroup#1\@@endlink}%
\providecommand \@sanitize@url [0]{\catcode `\\12\catcode `\$12\catcode
  `\&12\catcode `\#12\catcode `\^12\catcode `\_12\catcode `\%12\relax}%
\providecommand \@@startlink[1]{}%
\providecommand \@@endlink[0]{}%
\providecommand \url  [0]{\begingroup\@sanitize@url \@url }%
\providecommand \@url [1]{\endgroup\@href {#1}{\urlprefix }}%
\providecommand \urlprefix  [0]{URL }%
\providecommand \Eprint [0]{\href }%
\providecommand \doibase [0]{http://dx.doi.org/}%
\providecommand \selectlanguage [0]{\@gobble}%
\providecommand \bibinfo  [0]{\@secondoftwo}%
\providecommand \bibfield  [0]{\@secondoftwo}%
\providecommand \translation [1]{[#1]}%
\providecommand \BibitemOpen [0]{}%
\providecommand \bibitemStop [0]{}%
\providecommand \bibitemNoStop [0]{.\EOS\space}%
\providecommand \EOS [0]{\spacefactor3000\relax}%
\providecommand \BibitemShut  [1]{\csname bibitem#1\endcsname}%
\let\auto@bib@innerbib\@empty
\bibitem [{\citenamefont {Rosenband}\ \emph {et~al.}(2008)\citenamefont
  {Rosenband}, \citenamefont {Hume}, \citenamefont {Schmidt}, \citenamefont
  {Chou}, \citenamefont {Brusch}, \citenamefont {Lorini}, \citenamefont
  {Oskay}, \citenamefont {Drullinger}, \citenamefont {Fortier}, \citenamefont
  {Stalnaker}, \citenamefont {Diddams}, \citenamefont {Swann}, \citenamefont
  {Newbury}, \citenamefont {Itano}, \citenamefont {Wineland},\ and\
  \citenamefont {Bergquist}}]{Rosenband.2008}%
  \BibitemOpen
  \bibfield  {author} {\bibinfo {author} {\bibfnamefont {T.}~\bibnamefont
  {Rosenband}}, \bibinfo {author} {\bibfnamefont {D.~B.}\ \bibnamefont {Hume}},
  \bibinfo {author} {\bibfnamefont {P.~O.}\ \bibnamefont {Schmidt}}, \bibinfo
  {author} {\bibfnamefont {C.~W.}\ \bibnamefont {Chou}}, \bibinfo {author}
  {\bibfnamefont {A.}~\bibnamefont {Brusch}}, \bibinfo {author} {\bibfnamefont
  {L.}~\bibnamefont {Lorini}}, \bibinfo {author} {\bibfnamefont {W.~H.}\
  \bibnamefont {Oskay}}, \bibinfo {author} {\bibfnamefont {R.~E.}\ \bibnamefont
  {Drullinger}}, \bibinfo {author} {\bibfnamefont {T.~M.}\ \bibnamefont
  {Fortier}}, \bibinfo {author} {\bibfnamefont {J.~E.}\ \bibnamefont
  {Stalnaker}}, \bibinfo {author} {\bibfnamefont {S.~A.}\ \bibnamefont
  {Diddams}}, \bibinfo {author} {\bibfnamefont {W.~C.}\ \bibnamefont {Swann}},
  \bibinfo {author} {\bibfnamefont {N.~R.}\ \bibnamefont {Newbury}}, \bibinfo
  {author} {\bibfnamefont {W.~M.}\ \bibnamefont {Itano}}, \bibinfo {author}
  {\bibfnamefont {D.~J.}\ \bibnamefont {Wineland}}, \ and\ \bibinfo {author}
  {\bibfnamefont {J.~C.}\ \bibnamefont {Bergquist}},\ }\href {\doibase
  10.1126/science.1154622} {\bibfield  {journal} {\bibinfo  {journal}
  {Science}\ }\textbf {\bibinfo {volume} {319}},\ \bibinfo {pages} {1808}
  (\bibinfo {year} {2008})}\BibitemShut {NoStop}%
\bibitem [{\citenamefont {Leefer}\ \emph {et~al.}(2013)\citenamefont {Leefer},
  \citenamefont {Weber}, \citenamefont {Cing{\"o}z}, \citenamefont
  {Torgerson},\ and\ \citenamefont {Budker}}]{Leefer.2013}%
  \BibitemOpen
  \bibfield  {author} {\bibinfo {author} {\bibfnamefont {N.}~\bibnamefont
  {Leefer}}, \bibinfo {author} {\bibfnamefont {C.~T.~M.}\ \bibnamefont
  {Weber}}, \bibinfo {author} {\bibfnamefont {A.}~\bibnamefont {Cing{\"o}z}},
  \bibinfo {author} {\bibfnamefont {J.~R.}\ \bibnamefont {Torgerson}}, \ and\
  \bibinfo {author} {\bibfnamefont {D.}~\bibnamefont {Budker}},\ }\href
  {http://link.aps.org/doi/10.1103/PhysRevLett.111.060801} {\bibfield
  {journal} {\bibinfo  {journal} {Physical Review Letters}\ }\textbf {\bibinfo
  {volume} {111}},\ \bibinfo {pages} {060801} (\bibinfo {year}
  {2013})}\BibitemShut {NoStop}%
\bibitem [{\citenamefont {Cing{\"o}z}\ \emph {et~al.}(2007)\citenamefont
  {Cing{\"o}z}, \citenamefont {Lapierre}, \citenamefont {Nguyen}, \citenamefont
  {Leefer}, \citenamefont {Budker}, \citenamefont {Lamoreaux},\ and\
  \citenamefont {Torgerson}}]{Cingoz.2007}%
  \BibitemOpen
  \bibfield  {author} {\bibinfo {author} {\bibfnamefont {A.}~\bibnamefont
  {Cing{\"o}z}}, \bibinfo {author} {\bibfnamefont {A.}~\bibnamefont
  {Lapierre}}, \bibinfo {author} {\bibfnamefont {A.-T.}\ \bibnamefont
  {Nguyen}}, \bibinfo {author} {\bibfnamefont {N.}~\bibnamefont {Leefer}},
  \bibinfo {author} {\bibfnamefont {D.}~\bibnamefont {Budker}}, \bibinfo
  {author} {\bibfnamefont {S.~K.}\ \bibnamefont {Lamoreaux}}, \ and\ \bibinfo
  {author} {\bibfnamefont {J.~R.}\ \bibnamefont {Torgerson}},\ }\href {\doibase
  10.1103/PhysRevLett.98.040801} {\bibfield  {journal} {\bibinfo  {journal}
  {Phys. Rev. Let.}\ }\textbf {\bibinfo {volume} {98}},\ \bibinfo {pages}
  {040801} (\bibinfo {year} {2007})}\BibitemShut {NoStop}%
\bibitem [{\citenamefont {Ferrell}\ \emph {et~al.}(2007)\citenamefont
  {Ferrell}, \citenamefont {Cing{\"o}z}, \citenamefont {Lapierre},
  \citenamefont {Nguyen}, \citenamefont {Leefer}, \citenamefont {Budker},
  \citenamefont {Flambaum}, \citenamefont {Lamoreaux},\ and\ \citenamefont
  {Torgerson}}]{Ferrell.2007}%
  \BibitemOpen
  \bibfield  {author} {\bibinfo {author} {\bibfnamefont {S.~J.}\ \bibnamefont
  {Ferrell}}, \bibinfo {author} {\bibfnamefont {A.}~\bibnamefont {Cing{\"o}z}},
  \bibinfo {author} {\bibfnamefont {A.}~\bibnamefont {Lapierre}}, \bibinfo
  {author} {\bibfnamefont {A.-T.}\ \bibnamefont {Nguyen}}, \bibinfo {author}
  {\bibfnamefont {N.}~\bibnamefont {Leefer}}, \bibinfo {author} {\bibfnamefont
  {D.}~\bibnamefont {Budker}}, \bibinfo {author} {\bibfnamefont {V.~V.}\
  \bibnamefont {Flambaum}}, \bibinfo {author} {\bibfnamefont {S.~K.}\
  \bibnamefont {Lamoreaux}}, \ and\ \bibinfo {author} {\bibfnamefont {J.~R.}\
  \bibnamefont {Torgerson}},\ }\href {\doibase 10.1103/PhysRevA.76.062104}
  {\bibfield  {journal} {\bibinfo  {journal} {Phys. Rev. A}\ }\textbf {\bibinfo
  {volume} {76}},\ \bibinfo {pages} {062104} (\bibinfo {year}
  {2007})}\BibitemShut {NoStop}%
\bibitem [{\citenamefont {Gu{\'e}na}\ \emph {et~al.}(2012)\citenamefont
  {Gu{\'e}na}, \citenamefont {Abgrall}, \citenamefont {Rovera}, \citenamefont
  {Rosenbusch}, \citenamefont {Tobar}, \citenamefont {Laurent}, \citenamefont
  {Clairon},\ and\ \citenamefont {Bize}}]{Guena.2012}%
  \BibitemOpen
  \bibfield  {author} {\bibinfo {author} {\bibfnamefont {J.}~\bibnamefont
  {Gu{\'e}na}}, \bibinfo {author} {\bibfnamefont {M.}~\bibnamefont {Abgrall}},
  \bibinfo {author} {\bibfnamefont {D.}~\bibnamefont {Rovera}}, \bibinfo
  {author} {\bibfnamefont {P.}~\bibnamefont {Rosenbusch}}, \bibinfo {author}
  {\bibfnamefont {M.~E.}\ \bibnamefont {Tobar}}, \bibinfo {author}
  {\bibfnamefont {P.}~\bibnamefont {Laurent}}, \bibinfo {author} {\bibfnamefont
  {A.}~\bibnamefont {Clairon}}, \ and\ \bibinfo {author} {\bibfnamefont
  {S.}~\bibnamefont {Bize}},\ }\href {\doibase 10.1103/PhysRevLett.109.080801}
  {\bibfield  {journal} {\bibinfo  {journal} {Phys. Rev. Let.}\ }\textbf
  {\bibinfo {volume} {109}},\ \bibinfo {pages} {080801} (\bibinfo {year}
  {2012})}\BibitemShut {NoStop}%
\bibitem [{\citenamefont {{M. A. Hohensee}}\ \emph {et~al.}(2013)\citenamefont
  {{M. A. Hohensee}}, \citenamefont {Leefer}, \citenamefont {Budker},
  \citenamefont {Harabati}, \citenamefont {Dzuba},\ and\ \citenamefont
  {Flambaum}}]{HohenseeM.A..2013}%
  \BibitemOpen
  \bibfield  {author} {\bibinfo {author} {\bibnamefont {{M. A. Hohensee}}},
  \bibinfo {author} {\bibfnamefont {N.}~\bibnamefont {Leefer}}, \bibinfo
  {author} {\bibfnamefont {D.}~\bibnamefont {Budker}}, \bibinfo {author}
  {\bibfnamefont {C.}~\bibnamefont {Harabati}}, \bibinfo {author}
  {\bibfnamefont {V.~A.}\ \bibnamefont {Dzuba}}, \ and\ \bibinfo {author}
  {\bibfnamefont {V.~V.}\ \bibnamefont {Flambaum}},\ }\href
  {http://link.aps.org/doi/10.1103/PhysRevLett.111.050401} {\bibfield
  {journal} {\bibinfo  {journal} {Physical Review Letters}\ }\textbf {\bibinfo
  {volume} {111}},\ \bibinfo {pages} {050401} (\bibinfo {year}
  {2013})}\BibitemShut {NoStop}%
\bibitem [{\citenamefont {Altschul}(2010)}]{Altschul.2010}%
  \BibitemOpen
  \bibfield  {author} {\bibinfo {author} {\bibfnamefont {B.}~\bibnamefont
  {Altschul}},\ }\href {http://link.aps.org/doi/10.1103/PhysRevD.81.041701}
  {\bibfield  {journal} {\bibinfo  {journal} {Physical Review D}\ }\textbf
  {\bibinfo {volume} {81}},\ \bibinfo {pages} {041701} (\bibinfo {year}
  {2010})}\BibitemShut {NoStop}%
\bibitem [{\citenamefont {Dzuba}\ and\ \citenamefont
  {Flambaum}(2008)}]{Dzuba.2008}%
  \BibitemOpen
  \bibfield  {author} {\bibinfo {author} {\bibfnamefont {V.~A.}\ \bibnamefont
  {Dzuba}}\ and\ \bibinfo {author} {\bibfnamefont {V.~V.}\ \bibnamefont
  {Flambaum}},\ }\href {\doibase 10.1103/PhysRevA.77.012515} {\bibfield
  {journal} {\bibinfo  {journal} {Phys. Rev. A}\ }\textbf {\bibinfo {volume}
  {77}},\ \bibinfo {pages} {012515} (\bibinfo {year} {2008})}\BibitemShut
  {NoStop}%
\bibitem [{\citenamefont {Kosteleck\'{y}}\ and\ \citenamefont
  {Lane}(1999)}]{Kostelecky1999a}%
  \BibitemOpen
  \bibfield  {author} {\bibinfo {author} {\bibfnamefont {V.~A.}\ \bibnamefont
  {Kosteleck\'{y}}}\ and\ \bibinfo {author} {\bibfnamefont {C.~D.}\
  \bibnamefont {Lane}},\ }\href {\doibase 10.1103/PhysRevD.60.116010}
  {\bibfield  {journal} {\bibinfo  {journal} {Physical Review D}\ }\textbf
  {\bibinfo {volume} {60}},\ \bibinfo {pages} {116010} (\bibinfo {year}
  {1999})}\BibitemShut {NoStop}%
\bibitem [{\citenamefont {Budker}\ \emph {et~al.}(1994)\citenamefont {Budker},
  \citenamefont {DeMille}, \citenamefont {Commins},\ and\ \citenamefont
  {Zolotorev}}]{Budker.1994}%
  \BibitemOpen
  \bibfield  {author} {\bibinfo {author} {\bibfnamefont {D.}~\bibnamefont
  {Budker}}, \bibinfo {author} {\bibfnamefont {D.}~\bibnamefont {DeMille}},
  \bibinfo {author} {\bibfnamefont {E.~D.}\ \bibnamefont {Commins}}, \ and\
  \bibinfo {author} {\bibfnamefont {M.~S.}\ \bibnamefont {Zolotorev}},\ }\href
  {\doibase 10.1103/PhysRevA.50.132} {\bibfield  {journal} {\bibinfo  {journal}
  {Phys. Rev. A}\ }\textbf {\bibinfo {volume} {50}},\ \bibinfo {pages} {132}
  (\bibinfo {year} {1994})}\BibitemShut {NoStop}%
\bibitem [{\citenamefont {Dzuba}\ and\ \citenamefont
  {Flambaum}(2010)}]{Dzuba.2010}%
  \BibitemOpen
  \bibfield  {author} {\bibinfo {author} {\bibfnamefont {V.~A.}\ \bibnamefont
  {Dzuba}}\ and\ \bibinfo {author} {\bibfnamefont {V.~V.}\ \bibnamefont
  {Flambaum}},\ }\href {\doibase 10.1103/PhysRevA.81.052515} {\bibfield
  {journal} {\bibinfo  {journal} {Phys. Rev. A}\ }\textbf {\bibinfo {volume}
  {81}},\ \bibinfo {pages} {052515} (\bibinfo {year} {2010})}\BibitemShut
  {NoStop}%
\bibitem [{\citenamefont {{J. M. Amini}}\ and\ \citenamefont
  {Gould}(2003)}]{Amini.2003}%
  \BibitemOpen
  \bibfield  {author} {\bibinfo {author} {\bibnamefont {{J. M. Amini}}}\ and\
  \bibinfo {author} {\bibfnamefont {H.}~\bibnamefont {Gould}},\ }\href
  {\doibase 10.1103/PhysRevLett.91.153001} {\bibfield  {journal} {\bibinfo
  {journal} {Physical Review Letters}\ }\textbf {\bibinfo {volume} {91}},\
  \bibinfo {pages} {153001} (\bibinfo {year} {2003})}\BibitemShut {NoStop}%
\bibitem [{\citenamefont {Ulzega}\ \emph {et~al.}(2006)\citenamefont {Ulzega},
  \citenamefont {Hofer}, \citenamefont {Moroshkin},\ and\ \citenamefont
  {Weis}}]{Ulzega2006}%
  \BibitemOpen
  \bibfield  {author} {\bibinfo {author} {\bibfnamefont {S.}~\bibnamefont
  {Ulzega}}, \bibinfo {author} {\bibfnamefont {A.}~\bibnamefont {Hofer}},
  \bibinfo {author} {\bibfnamefont {P.}~\bibnamefont {Moroshkin}}, \ and\
  \bibinfo {author} {\bibfnamefont {A.}~\bibnamefont {Weis}},\ }\href {\doibase
  10.1209/epl/i2006-10383-2} {\bibfield  {journal} {\bibinfo  {journal}
  {Europhysics Letters (EPL)}\ }\textbf {\bibinfo {volume} {76}},\ \bibinfo
  {pages} {1074} (\bibinfo {year} {2006})}\BibitemShut {NoStop}%
\bibitem [{\citenamefont {Itano}\ \emph {et~al.}(1982)\citenamefont {Itano},
  \citenamefont {Lewis},\ and\ \citenamefont {Wineland}}]{Itano.1982}%
  \BibitemOpen
  \bibfield  {author} {\bibinfo {author} {\bibfnamefont {W.~M.}\ \bibnamefont
  {Itano}}, \bibinfo {author} {\bibfnamefont {L.~L.}\ \bibnamefont {Lewis}}, \
  and\ \bibinfo {author} {\bibfnamefont {D.~J.}\ \bibnamefont {Wineland}},\
  }\href {\doibase 10.1103/PhysRevA.25.1233} {\bibfield  {journal} {\bibinfo
  {journal} {Phys. Rev. A}\ }\textbf {\bibinfo {volume} {25}},\ \bibinfo
  {pages} {1233} (\bibinfo {year} {1982})}\BibitemShut {NoStop}%
\bibitem [{\citenamefont {Levi}\ \emph {et~al.}(2004)\citenamefont {Levi},
  \citenamefont {Calonico}, \citenamefont {Lorini}, \citenamefont {Micalizio},\
  and\ \citenamefont {Godone}}]{Levi.2004}%
  \BibitemOpen
  \bibfield  {author} {\bibinfo {author} {\bibfnamefont {F.}~\bibnamefont
  {Levi}}, \bibinfo {author} {\bibfnamefont {D.}~\bibnamefont {Calonico}},
  \bibinfo {author} {\bibfnamefont {L.}~\bibnamefont {Lorini}}, \bibinfo
  {author} {\bibfnamefont {S.}~\bibnamefont {Micalizio}}, \ and\ \bibinfo
  {author} {\bibfnamefont {A.}~\bibnamefont {Godone}},\ }\href {\doibase
  10.1103/PhysRevA.70.033412} {\bibfield  {journal} {\bibinfo  {journal} {Phys.
  Rev. A}\ }\textbf {\bibinfo {volume} {70}},\ \bibinfo {pages} {033412}
  (\bibinfo {year} {2004})}\BibitemShut {NoStop}%
\bibitem [{\citenamefont {Nguyen}\ \emph {et~al.}(2004)\citenamefont {Nguyen},
  \citenamefont {Budker}, \citenamefont {Lamoreaux},\ and\ \citenamefont
  {Torgerson}}]{Nguyen.2004}%
  \BibitemOpen
  \bibfield  {author} {\bibinfo {author} {\bibfnamefont {A.~T.}\ \bibnamefont
  {Nguyen}}, \bibinfo {author} {\bibfnamefont {D.}~\bibnamefont {Budker}},
  \bibinfo {author} {\bibfnamefont {S.~K.}\ \bibnamefont {Lamoreaux}}, \ and\
  \bibinfo {author} {\bibfnamefont {J.~R.}\ \bibnamefont {Torgerson}},\ }\href
  {\doibase 10.1103/PhysRevA.69.022105} {\bibfield  {journal} {\bibinfo
  {journal} {Phys. Rev. A}\ }\textbf {\bibinfo {volume} {69}},\ \bibinfo
  {pages} {022105} (\bibinfo {year} {2004})}\BibitemShut {NoStop}%
\bibitem [{\citenamefont {Bonin}\ and\ \citenamefont
  {Kadar-Kallen}(1994)}]{Bonin1994}%
  \BibitemOpen
  \bibfield  {author} {\bibinfo {author} {\bibfnamefont {K.~D.}\ \bibnamefont
  {Bonin}}\ and\ \bibinfo {author} {\bibfnamefont {M.~A.}\ \bibnamefont
  {Kadar-Kallen}},\ }\href {\doibase 10.1142/S0217979294001391} {\bibfield
  {journal} {\bibinfo  {journal} {International Journal of Modern Physics B}\
  }\textbf {\bibinfo {volume} {08}},\ \bibinfo {pages} {3313} (\bibinfo {year}
  {1994})}\BibitemShut {NoStop}%
\bibitem [{\citenamefont {Nguyen}\ \emph {et~al.}(1997)\citenamefont {Nguyen},
  \citenamefont {Budker}, \citenamefont {DeMille},\ and\ \citenamefont
  {Zolotorev}}]{Nguyen.1997}%
  \BibitemOpen
  \bibfield  {author} {\bibinfo {author} {\bibfnamefont {A.~T.}\ \bibnamefont
  {Nguyen}}, \bibinfo {author} {\bibfnamefont {D.}~\bibnamefont {Budker}},
  \bibinfo {author} {\bibfnamefont {D.}~\bibnamefont {DeMille}}, \ and\
  \bibinfo {author} {\bibfnamefont {M.}~\bibnamefont {Zolotorev}},\ }\href
  {\doibase 10.1103/PhysRevA.56.3453} {\bibfield  {journal} {\bibinfo
  {journal} {Phys. Rev. A}\ }\textbf {\bibinfo {volume} {56}},\ \bibinfo
  {pages} {3453} (\bibinfo {year} {1997})}\BibitemShut {NoStop}%
\bibitem [{\citenamefont {Leefer}\ \emph {et~al.}(2010)\citenamefont {Leefer},
  \citenamefont {Cing{\"o}z}, \citenamefont {Gerber-Siff}, \citenamefont
  {Sharma}, \citenamefont {Torgerson},\ and\ \citenamefont
  {Budker}}]{Leefer.2010}%
  \BibitemOpen
  \bibfield  {author} {\bibinfo {author} {\bibfnamefont {N.}~\bibnamefont
  {Leefer}}, \bibinfo {author} {\bibfnamefont {A.}~\bibnamefont {Cing{\"o}z}},
  \bibinfo {author} {\bibfnamefont {B.}~\bibnamefont {Gerber-Siff}}, \bibinfo
  {author} {\bibfnamefont {A.}~\bibnamefont {Sharma}}, \bibinfo {author}
  {\bibfnamefont {J.~R.}\ \bibnamefont {Torgerson}}, \ and\ \bibinfo {author}
  {\bibfnamefont {D.}~\bibnamefont {Budker}},\ }\href {\doibase
  10.1103/PhysRevA.81.043427} {\bibfield  {journal} {\bibinfo  {journal} {Phys.
  Rev. A}\ }\textbf {\bibinfo {volume} {81}},\ \bibinfo {pages} {043427}
  (\bibinfo {year} {2010})}\BibitemShut {NoStop}%
\bibitem [{\citenamefont {Lu}\ \emph {et~al.}(2010)\citenamefont {Lu},
  \citenamefont {Youn},\ and\ \citenamefont {Lev}}]{Lu.2010}%
  \BibitemOpen
  \bibfield  {author} {\bibinfo {author} {\bibfnamefont {M.}~\bibnamefont
  {Lu}}, \bibinfo {author} {\bibfnamefont {S.~H.}\ \bibnamefont {Youn}}, \ and\
  \bibinfo {author} {\bibfnamefont {B.~L.}\ \bibnamefont {Lev}},\ }\href
  {\doibase 10.1103/PhysRevLett.104.063001} {\bibfield  {journal} {\bibinfo
  {journal} {Phys. Rev. Let.}\ }\textbf {\bibinfo {volume} {104}},\ \bibinfo
  {pages} {063001} (\bibinfo {year} {2010})}\BibitemShut {NoStop}%
\bibitem [{\citenamefont {Lu}\ \emph {et~al.}(2011)\citenamefont {Lu},
  \citenamefont {{N. Q. Burdick}}, \citenamefont {{S. H. Youn}},\ and\
  \citenamefont {{B. L. Lev}}}]{Lu.2011}%
  \BibitemOpen
  \bibfield  {author} {\bibinfo {author} {\bibfnamefont {M.}~\bibnamefont
  {Lu}}, \bibinfo {author} {\bibnamefont {{N. Q. Burdick}}}, \bibinfo {author}
  {\bibnamefont {{S. H. Youn}}}, \ and\ \bibinfo {author} {\bibnamefont {{B. L.
  Lev}}},\ }\href {\doibase 10.1103/PhysRevLett.107.190401} {\bibfield
  {journal} {\bibinfo  {journal} {Phys. Rev. Let.}\ }\textbf {\bibinfo {volume}
  {107}},\ \bibinfo {pages} {190401} (\bibinfo {year} {2011})}\BibitemShut
  {NoStop}%
\bibitem [{\citenamefont {Lu}\ \emph {et~al.}(2012)\citenamefont {Lu},
  \citenamefont {{N. Q. Burdick}},\ and\ \citenamefont {{B. L.
  Lev}}}]{Lu.2012}%
  \BibitemOpen
  \bibfield  {author} {\bibinfo {author} {\bibfnamefont {M.}~\bibnamefont
  {Lu}}, \bibinfo {author} {\bibnamefont {{N. Q. Burdick}}}, \ and\ \bibinfo
  {author} {\bibnamefont {{B. L. Lev}}},\ }\href {\doibase
  10.1103/PhysRevLett.108.215301} {\bibfield  {journal} {\bibinfo  {journal}
  {Phys. Rev. Let.}\ }\textbf {\bibinfo {volume} {108}},\ \bibinfo {pages}
  {215301} (\bibinfo {year} {2012})}\BibitemShut {NoStop}%
\bibitem [{\citenamefont {Derevianko}\ and\ \citenamefont
  {Cannon}(2004)}]{Derevianko.2004}%
  \BibitemOpen
  \bibfield  {author} {\bibinfo {author} {\bibfnamefont {A.}~\bibnamefont
  {Derevianko}}\ and\ \bibinfo {author} {\bibfnamefont {C.~C.}\ \bibnamefont
  {Cannon}},\ }\href {http://link.aps.org/doi/10.1103/PhysRevA.70.062319}
  {\bibfield  {journal} {\bibinfo  {journal} {Phys. Rev. A}\ }\textbf {\bibinfo
  {volume} {70}},\ \bibinfo {pages} {062319} (\bibinfo {year}
  {2004})}\BibitemShut {NoStop}%
\bibitem [{\citenamefont {Fregoso}\ \emph {et~al.}(2009)\citenamefont
  {Fregoso}, \citenamefont {Sun}, \citenamefont {Fradkin},\ and\ \citenamefont
  {Lev}}]{Fregoso.2009}%
  \BibitemOpen
  \bibfield  {author} {\bibinfo {author} {\bibfnamefont {B.~M.}\ \bibnamefont
  {Fregoso}}, \bibinfo {author} {\bibfnamefont {K.}~\bibnamefont {Sun}},
  \bibinfo {author} {\bibfnamefont {E.}~\bibnamefont {Fradkin}}, \ and\
  \bibinfo {author} {\bibfnamefont {B.~L.}\ \bibnamefont {Lev}},\ }\href
  {\doibase 10.1088/1367-2630/11/10/103003} {\bibfield  {journal} {\bibinfo
  {journal} {New Journal of Physics}\ }\textbf {\bibinfo {volume} {11}},\
  \bibinfo {pages} {103003} (\bibinfo {year} {2009})}\BibitemShut {NoStop}%
\bibitem [{\citenamefont {{Jean-Philippe Uzan}}(2011)}]{JeanPhilippeUzan.2011}%
  \BibitemOpen
  \bibfield  {author} {\bibinfo {author} {\bibnamefont {{Jean-Philippe
  Uzan}}},\ }\href {http://www.livingreviews.org/lrr-2011-2} {\bibfield
  {journal} {\bibinfo  {journal} {Living Reviews in Relativity}\ }\textbf
  {\bibinfo {volume} {14}} (\bibinfo {year} {2011})}\BibitemShut {NoStop}%
\bibitem [{\citenamefont {{David Mattingly}}(2005)}]{DavidMattingly.2005}%
  \BibitemOpen
  \bibfield  {author} {\bibinfo {author} {\bibnamefont {{David Mattingly}}},\
  }\href {http://www.livingreviews.org/lrr-2005-5 \% \% Note: Encoded in UTF-8!
  To use with TeX, adapt the encoding \% (with tools ’recode’ or
  ’iconv’) or use the ucs package \%
  (http://ctan.tug.org/tex-archive/macros/latex/contrib/unicode/) \%}
  {\bibfield  {journal} {\bibinfo  {journal} {Living Reviews in Relativity}\
  }\textbf {\bibinfo {volume} {8}} (\bibinfo {year} {2005})}\BibitemShut
  {NoStop}%
\bibitem [{\citenamefont {Flambaum}\ and\ \citenamefont
  {Dzuba}(2009)}]{Flambaum.2009b}%
  \BibitemOpen
  \bibfield  {author} {\bibinfo {author} {\bibfnamefont {V.~V.}\ \bibnamefont
  {Flambaum}}\ and\ \bibinfo {author} {\bibfnamefont {V.~A.}\ \bibnamefont
  {Dzuba}},\ }\href {\doibase 10.1139/p08-072} {\bibfield  {journal} {\bibinfo
  {journal} {Can. J. Phys}\ }\textbf {\bibinfo {volume} {87}},\ \bibinfo
  {pages} {25} (\bibinfo {year} {2009})}\BibitemShut {NoStop}%
\bibitem [{\citenamefont {Geremia}\ \emph {et~al.}(2006)\citenamefont
  {Geremia}, \citenamefont {Stockton},\ and\ \citenamefont
  {Mabuchi}}]{Geremia.2006}%
  \BibitemOpen
  \bibfield  {author} {\bibinfo {author} {\bibfnamefont {J.~M.}\ \bibnamefont
  {Geremia}}, \bibinfo {author} {\bibfnamefont {J.~K.}\ \bibnamefont
  {Stockton}}, \ and\ \bibinfo {author} {\bibfnamefont {H.}~\bibnamefont
  {Mabuchi}},\ }\href {http://link.aps.org/doi/10.1103/PhysRevA.73.042112}
  {\bibfield  {journal} {\bibinfo  {journal} {Physical Review A}\ }\textbf
  {\bibinfo {volume} {73}},\ \bibinfo {pages} {042112} (\bibinfo {year}
  {2006})}\BibitemShut {NoStop}%
\bibitem [{\citenamefont {Dounas-Frazer}\ \emph {et~al.}(2010)\citenamefont
  {Dounas-Frazer}, \citenamefont {Tsigutkin}, \citenamefont {Family},\ and\
  \citenamefont {Budker}}]{DounasFrazer.2010}%
  \BibitemOpen
  \bibfield  {author} {\bibinfo {author} {\bibfnamefont {D.~R.}\ \bibnamefont
  {Dounas-Frazer}}, \bibinfo {author} {\bibfnamefont {K.}~\bibnamefont
  {Tsigutkin}}, \bibinfo {author} {\bibfnamefont {A.}~\bibnamefont {Family}}, \
  and\ \bibinfo {author} {\bibfnamefont {D.}~\bibnamefont {Budker}},\ }\href
  {http://link.aps.org/doi/10.1103/PhysRevA.82.062507} {\bibfield  {journal}
  {\bibinfo  {journal} {Phys. Rev. A}\ }\textbf {\bibinfo {volume} {82}},\
  \bibinfo {pages} {062507} (\bibinfo {year} {2010})}\BibitemShut {NoStop}%
\bibitem [{\citenamefont {Auzinsh}\ \emph {et~al.}(2010)\citenamefont
  {Auzinsh}, \citenamefont {Budker},\ and\ \citenamefont
  {Rochester}}]{Auzinsh.2010}%
  \BibitemOpen
  \bibfield  {author} {\bibinfo {author} {\bibfnamefont {M.}~\bibnamefont
  {Auzinsh}}, \bibinfo {author} {\bibfnamefont {D.}~\bibnamefont {Budker}}, \
  and\ \bibinfo {author} {\bibfnamefont {S.~M.}\ \bibnamefont {Rochester}},\
  }\href@noop {} {\emph {\bibinfo {title} {Optically Polarized Atoms:
  Understanding Light-Atom Interactions}}}\ (\bibinfo  {publisher} {Oxford
  University Press, USA},\ \bibinfo {year} {2010})\BibitemShut {NoStop}%
\bibitem [{\citenamefont {Budker}\ \emph {et~al.}(2008)\citenamefont {Budker},
  \citenamefont {Kimball},\ and\ \citenamefont {DeMille}}]{Budker.2008}%
  \BibitemOpen
  \bibfield  {author} {\bibinfo {author} {\bibfnamefont {D.}~\bibnamefont
  {Budker}}, \bibinfo {author} {\bibfnamefont {D.}~\bibnamefont {Kimball}}, \
  and\ \bibinfo {author} {\bibfnamefont {D.}~\bibnamefont {DeMille}},\
  }\href@noop {} {\emph {\bibinfo {title} {Atomic physics: An exploration
  through problems and solutions: Second Edition}}}\ (\bibinfo  {publisher}
  {Oxford University Press, USA},\ \bibinfo {year} {2008})\BibitemShut
  {NoStop}%
\bibitem [{\citenamefont {{Benjamin J.
  Sussman}}(2011)}]{BenjaminJ.Sussman.2011}%
  \BibitemOpen
  \bibfield  {author} {\bibinfo {author} {\bibnamefont {{Benjamin J.
  Sussman}}},\ }\href {http://link.aip.org/link/?AJP/79/477/1} {\bibfield
  {journal} {\bibinfo  {journal} {American Journal of Physics}\ }\textbf
  {\bibinfo {volume} {79}},\ \bibinfo {pages} {477} (\bibinfo {year}
  {2011})}\BibitemShut {NoStop}%
\bibitem [{\citenamefont {Kramida}\ \emph {et~al.}(2012)\citenamefont
  {Kramida}, \citenamefont {Ralchenko},\ and\ \citenamefont {{Reader J. and
  NIST ASD Team}}}]{kramida.2012}%
  \BibitemOpen
  \bibfield  {author} {\bibinfo {author} {\bibfnamefont {A.}~\bibnamefont
  {Kramida}}, \bibinfo {author} {\bibfnamefont {Y.}~\bibnamefont {Ralchenko}},
  \ and\ \bibinfo {author} {\bibnamefont {{Reader J. and NIST ASD Team}}},\
  }\href {http://physics.nist.gov/asd} {\enquote {\bibinfo {title} {Nist atomic
  spectra database (ver. 5.0), [online].}}\ } (\bibinfo {year}
  {2012})\BibitemShut {NoStop}%
\bibitem [{\citenamefont {Cing{\"o}z}\ \emph {et~al.}(2005)\citenamefont
  {Cing{\"o}z}, \citenamefont {Nguyen}, \citenamefont {Budker}, \citenamefont
  {Lamoreaux}, \citenamefont {Lapierre},\ and\ \citenamefont
  {Torgerson}}]{Cingoz.2005}%
  \BibitemOpen
  \bibfield  {author} {\bibinfo {author} {\bibfnamefont {A.}~\bibnamefont
  {Cing{\"o}z}}, \bibinfo {author} {\bibfnamefont {A.-T.}\ \bibnamefont
  {Nguyen}}, \bibinfo {author} {\bibfnamefont {D.}~\bibnamefont {Budker}},
  \bibinfo {author} {\bibfnamefont {S.~K.}\ \bibnamefont {Lamoreaux}}, \bibinfo
  {author} {\bibfnamefont {A.}~\bibnamefont {Lapierre}}, \ and\ \bibinfo
  {author} {\bibfnamefont {J.~R.}\ \bibnamefont {Torgerson}},\ }\href
  {http://link.aps.org/doi/10.1103/PhysRevA.72.063409} {\bibfield  {journal}
  {\bibinfo  {journal} {Phys. Rev. A}\ }\textbf {\bibinfo {volume} {72}},\
  \bibinfo {pages} {063409} (\bibinfo {year} {2005})}\BibitemShut {NoStop}%
\bibitem [{\citenamefont {Nguyen}\ \emph {et~al.}(2000)\citenamefont {Nguyen},
  \citenamefont {Chern}, \citenamefont {Budker},\ and\ \citenamefont
  {Zolotorev}}]{Nguyen.2000}%
  \BibitemOpen
  \bibfield  {author} {\bibinfo {author} {\bibfnamefont {A.~T.}\ \bibnamefont
  {Nguyen}}, \bibinfo {author} {\bibfnamefont {G.~D.}\ \bibnamefont {Chern}},
  \bibinfo {author} {\bibfnamefont {D.}~\bibnamefont {Budker}}, \ and\ \bibinfo
  {author} {\bibfnamefont {M.}~\bibnamefont {Zolotorev}},\ }\href
  {http://link.aps.org/doi/10.1103/PhysRevA.63.013406} {\bibfield  {journal}
  {\bibinfo  {journal} {Phys. Rev. A}\ }\textbf {\bibinfo {volume} {63}},\
  \bibinfo {pages} {013406} (\bibinfo {year} {2000})}\BibitemShut {NoStop}%
\end{thebibliography}%

\end{document}